\documentclass{aastex63}

\received{February 26, 2021}
\revised{\today}
\accepted{September 9, 2021}
\submitjournal{PSJ}

\shorttitle{Young Alluvial Fans on Mars}
\shortauthors{Holo et al.}
\graphicspath{{./}{figures/}}

\begin{document}

\title{The Timing of Alluvial Fan Formation on Mars}

\correspondingauthor{Samuel Holo}
\email{samueljholo@gmail.com, kite@uchicago.edu}

\author[0000-0003-3232-7334]{Samuel J. Holo}
\affiliation{Department of the Geophysical Sciences \\
University of Chicago \\
Chicago, IL, USA}

\author[0000-0002-1426-1186]{Edwin S. Kite}
\affiliation{Department of the Geophysical Sciences \\
University of Chicago \\
Chicago, IL, USA}

\nocollaboration{2}

\author[0000-0001-8872-8546]{Sharon A. Wilson}
\affiliation{Center for Earth and Planetary Studies \\
National Air and Space Museum \\
Smithsonian Institution \\
Washington, DC, USA}

\author[0000-0003-2443-1676]{Alexander M. Morgan}
\affiliation{Center for Earth and Planetary Studies \\
National Air and Space Museum \\
Smithsonian Institution \\
Washington, DC, USA}
\affiliation{Planetary Science Institute \\
Tucson, AZ, USA}



\begin{abstract}

The history of rivers on Mars is an important constraint on Martian climate evolution. The timing of relatively young, alluvial fan-forming rivers is especially important, as Mars’ Amazonian atmosphere is thought to have been too thin to consistently support surface liquid water. Previous regional studies suggested that alluvial fans formed primarily between the Early Hesperian and the Early Amazonian. In this study, we describe how a combination of a global impact crater database, a global geologic map, a global alluvial fan database, and statistical models can be used to estimate the timing of alluvial fan formation across Mars. Using our global approach and improved statistical modeling, we find that alluvial fan formation likely persisted into the last $\sim 2.5$ Gyr, well into the Amazonian period. However, the data we analyzed was insufficient to place constraints on the duration of alluvial fan formation. Going forward, more crater data will enable tighter constraints on the parameters estimated in our models and thus further inform our understanding of Mars’ climate evolution.

\end{abstract}



\section{Introduction}

River deposits on Mars record past climatic conditions and thus serve as a proxy for Mars’ climate evolution towards its current cold and dry state \citep[e.g.][]{KITE2019}. Since the end of the era of regionally-integrated valley networks in the early Hesperian $\sim3.4$ Ga \citep{FASSETT2008}, fluvial activity on Mars primarily produced shorter landforms \citep{GOUDGE2016}, including small, ‘pollywog’, exit breach craters \citep{WILSON2016,WARREN2021}, alluvial fans \citep{MOORE2005,KRAAL2008,GRANT2011,HAUBER2013,KITE2017,GRANT2019,WILSON2021}, and gullies (which may or may not be fluvial in origin; see \citet{CONWAY2019}). Some of the alluvial fans are associated with aqueous minerals (hydrated silica; e.g. \citet{PAN2020}). The apparently recent ages of these fluvial features are at odds with the belief that intense atmospheric loss early in Mars’ history would have rapidly rendered Mars’ surface cold and dry \citep[e.g.][]{KITEETAL2019}. Further, Amazonian/Hesperian catastrophic floods \citep[e.g.][]{BAKER1974} are commonly (but not exclusively) believed to require a $\geq1$ km thick cryosphere (and thereby a cold climate) in order to over-pressurize the aquifers that fed these floods \citep[e.g.][]{CARR1979}. Multiple mechanisms have been proposed to explain late-forming rivers, including impact-induced snowmelt/precipitation \citep[e.g.][]{WILLIAMS2008,MANGOLD2012}, transient greenhouse atmospheres \citep[e.g.][]{KITE2021, WORDSWORTH2021}, and obliquity-shift induced snowmelt/precipitation \citep[e.g.][]{IRWIN2015}. Each of these scenarios predicts transient excursions from an otherwise cold, dry climate state. Different proposed mechanisms have different dependencies on volcanism and/or $p$CO$_2$, both of which declined over time \citep{FASSETT2011}. Thus, the timing of young fluvial activity is an important constraint on Martian climate evolution. 

In this study, we focus on the alluvial fans (for the purposes of this study, we also include deltas, which are fans formed in standing bodies of water) and how improved statistical methodologies shed light on their formation ages. The alluvial fans are of particular interest because many are precipitation-fed \citep{KITEETAL2019}, at least some formed over an extended time span ($> 20 - 200$ Myr) \citep{KITE2017}, and some are up to $\sim 1$ km thick, recording a minimum of 100 years – 1 Myr of intermittent river flow \citep{IRWIN2015,GAIA2019}. These observations, in addition to the widespread distribution of alluvial fans within observed latitude bands \citep{WILSON2021}, are consistent with a synoptic water source rather than localized impact trigger and therefore provide true constraints on the Martian climate through time. 

 The primary method for estimating the absolute age of geologic surfaces on Mars is through counting the craters in an area and comparing the observed density to a chronology function, which relates crater density to absolute age \citep[e.g.][and references therein]{MICHAEL2013,FASSETT2016}. Dating small areas like fan surfaces requires counting small craters which are, by virtue of their size, more susceptible to survey complications like atmospheric filtering \citep[e.g.][]{KITE2014} and geologic modification (e.g. erosion, burial; \citet{PALUCIS2020}). Indeed, derived ages for small areas such as individual deltas vary over a huge time range \citep[e.g.][]{HAUBER2013,WARNER2015}. To reduce these uncertainties, previous workers have averaged over numerous close-proximity fans \citep[e.g.][]{MOORE2005,GRANT2011,MORGAN2019}. However, this methodology (1) implicitly assumes that fan formation occurred simultaneously for each fan (such that their surfaces reflect a single age), (2) requires that there has not been sufficient, potentially spatially-varying, resurfacing of the fan surfaces to obscure any craters in the considered size range \citep{PALUCIS2020}, and (3) sets aside the likely possibility that alluvial fans formed intermittently over tens or hundreds of Myr, rather than instantaneously \citep{KITE2017,GAIA2019}. We propose that these shortcomings can be vaulted by a new approach involving careful consideration of the fans’ cross-cutting relationships and choice of statistical modeling methodology. 
 
 The majority of alluvial fans are hosted on the interior walls of impact craters \citep{WILSON2021}, where erosional backwasting of crater rims provided alluvium that was transported down-slope and deposited on crater floors. These fans unambiguously post-date their host-craters, whereas determination of alluvial fan stratigraphic relationships within host crater interiors requires more detailed investigation on an individual fan basis. Further, the ages of the impact craters can be estimated via crater counts on their ejecta deposits (where preserved), which provide us order-of-magnitude greater count areas (and thus the ability to count larger craters) than the fan surfaces themselves. In the following sections, we demonstrate how combining a global database of impact craters \citep{ROBBINS2012}, a global geologic map \citep{TANAKA2014}, a global database of alluvial fans \citep{WILSON2021}, and a novel statistical model can constrain alluvial fan formation timing to have persisted until $< 3$ Ga. Further, we will discuss limitations of our approach and how updated datasets will affect fan-timing inferences. 
 
 \section{Datasets and Methods}
 
 To investigate the formation of alluvial fans, we used a recently compiled global database of alluvial fans (including potential deltas) in craters \citep{WILSON2021} that incorporates previously published surveys \citep{MOORE2005,KRAAL2008} and abstracts  \citep[e.g.][]{MORGAN2019} in addition to newly identified fans using images from the Context Camera (CTX) \citep{MALIN2007}. Although the use of CTX imagery enabled surveying smaller fans than previous studies, we expect that the smallest features are missing from the survey, due to the finite resolution of the images used. For this study, we restricted the fans dataset to those contained within the 212 Amazonian-Hesperian Impact Unit (AHi) subunits from the \citet{TANAKA2014} global geologic map of Mars that are fully contained within $40^{\circ}$ latitude of the equator (Appendix Figure 1). These subunits are impact craters that have well-preserved rims and ejecta \citep{TANAKA2014}.These AHi subunits provide us a sample of areas that are globally distributed spatially, share similar steep slopes and topography, and record their ages via crater-counts on their ejecta. Because fans on the crater walls must post-date the crater formation, ages obtained from the ejecta provide robust upper bounds on each fan’s age. Further, not all AHi subunits possess fans (70 out of 212 subunits contain fans), which enables comparison between the formation times of fan-bearing and non-fan-bearing surfaces. 
 
\begin{figure}[hbt!]
\plottwo{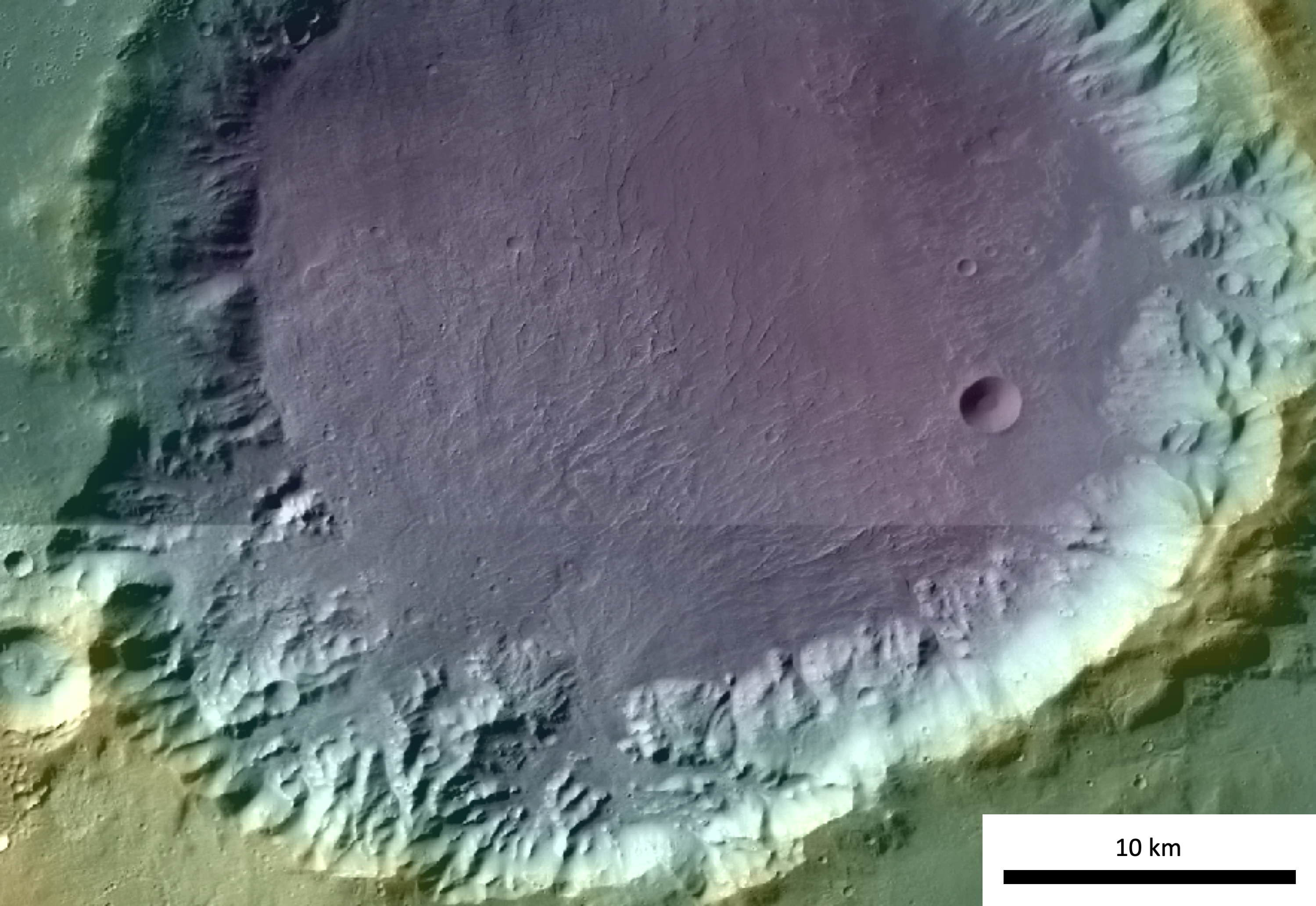}{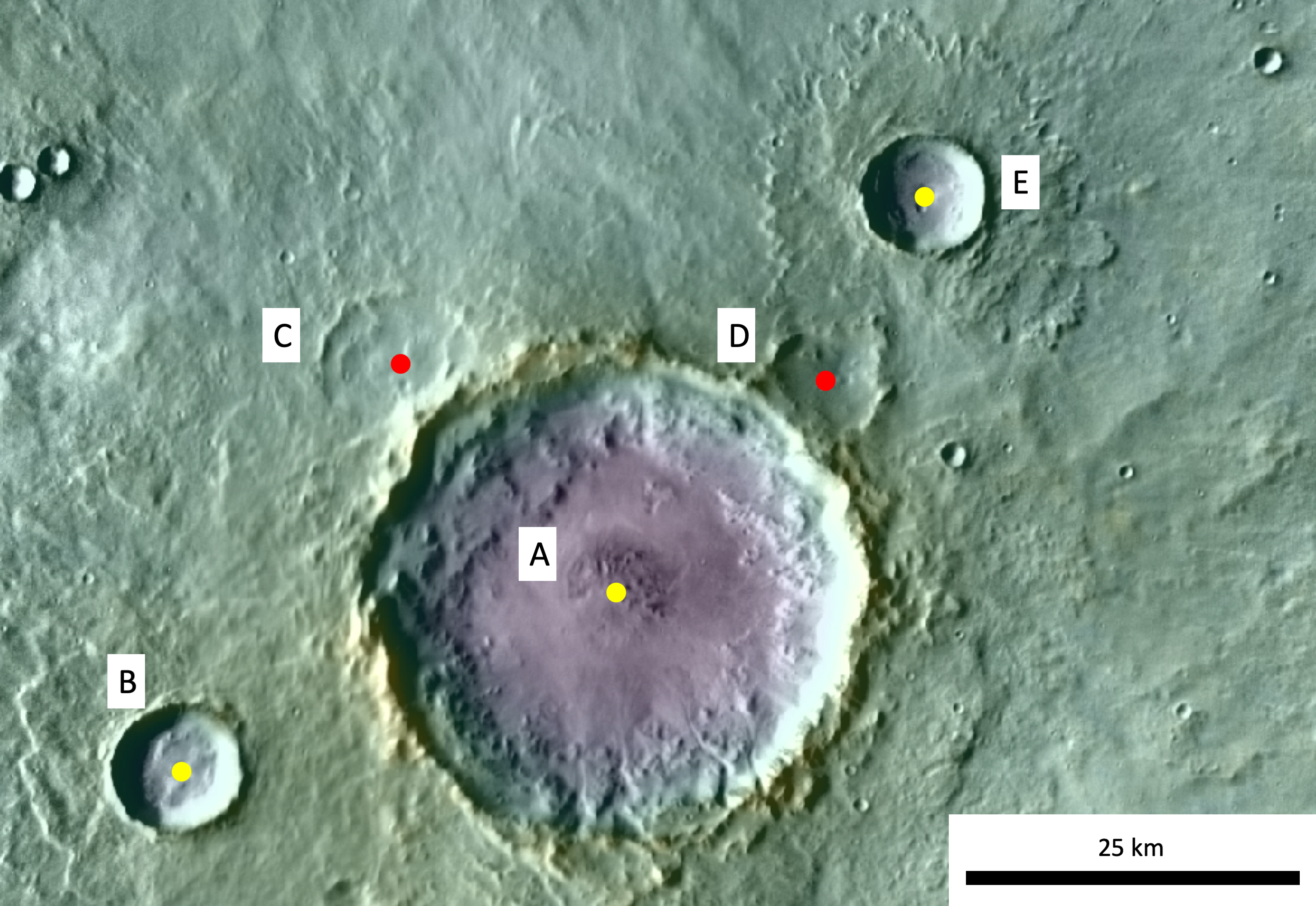}
\caption{Left: Colorized elevation, ranging from $\sim -1.1$ km (indigo) to $~\sim 1$ km (yellow), over Global CTX Mosaic
\citep{DICKSON2018} (illumination from north-west to south-east) image of a large crater-hosted alluvial fan deposit, formed by erosional backwasting of the southwestern crater rim $(333^\circ$ E$, 28^\circ $ S$)$. Right: Colorized elevation, ranging from $\sim -1.8$ km (indigo) to $~\sim 0.7$ km (yellow), over THEMIS daytime IR image of an AHi subunit $(0^\circ$ E$, 15^\circ$ S$)$ (illumination from west to east). In this example two craters (C, D) pre-date the central crater (A) based on cross-cutting relationships with the central crater’s northern rim and ejecta deposits. The crater in the top right (E) has well-preserved ejecta that cross-cuts the central crater’s ejecta deposits, indicating that it post-dated the AHi subunit formation. The crater in the bottom left (B) appears fresh and has ambiguous relationships with the terrain around it, so it was included in the count. In both images, blended High Resolution Stereo Camera (HRSC) and Mars Orbiter Laser Altimeter (MOLA) 200m elevation data \citep{FERGASON2018} was used to produce colorized elevation. In both images, north is towards the top of the page.}
\label{fig:fig1}
\end{figure}

 To obtain crater densities for each AHi subunits’s ejecta, we used a global database of impact craters $> 1$ km in diameter \citep{ROBBINS2012}. However, due to concerns over survey completeness for small craters (Stuart Robbins, personal communication), we restricted the dataset to craters with diameters of at least 4 km. Manual inspection of craters on the AHi subunits revealed three additional potential data quality issues. First, ejecta blankets often contain easily-identified chains of secondary craters. Second, some craters mapped on ejecta actually pre-date ejecta, as evidenced by cross-cutting relationships with the central crater’s rim, impact ejecta, or secondary craters (Figure \ref{fig:fig1}). Finally, craters in the interior of AHi subunits’ central craters are often obscured from surveys due to resurfacing by burial/erosion. 
 
 To remedy these crater-counting issues, we manually inspected each crater in Thermal Emission Imaging System (THEMIS) \citep{CHRISTENSEN2004} and CTX imagery and removed obvious secondary craters (elongated craters often in chains extending from the center of large nearby impact craters) and craters that obviously pre-date crater ejecta based on cross-cutting relationships. In general, we were conservative and did not discard ambiguous craters, preventing us from erroneously under-estimating the age of the subunits. This procedure, along with removing craters within each AHi subunits’s central crater’s interior, reduced our total crater count from 1768 to 1308. Further, we restricted our count areas on each AHi subunit to the annulus extending from the crater rim to 2$\times$ the subunit-forming-crater’s radius (i.e., we restricted our count areas to 1-2 crater radii from the crater center). In the $< 5\%$ of cases where the AHi subunit is defined by the ejecta blankets of 2 large craters, we chose the larger crater as our nominal center. In many cases, the ejecta blankets do not extend to 2$\times$ the crater radius, in which case we included in our count area only the intersection of the ejecta blanket and our defined annulus. We imposed the outer limit on our count area to accommodate the fact that craters “poking through” (i.e. pre-dating ejecta) near the ejecta edges cause crater densities to increase by a factor of $\sim2$--$3$ but that unambiguous cross-cutting relationships are increasingly difficult to identify as the ejecta thins radially outward. This indicates that the fraction of craters “poking-through” the ejecta blankets increases near the edges, and thus, the outer limit prevents erroneously biasing our AHi subunit ages upward. Imposing the outer annulus limit further reduced our total crater count to 498.
 
\section{Statistical Modeling and Results}

Our goal was to take crater counts on the AHi subunits’ ejecta blankets, note which subunits host alluvial fans, and make quantitative statements about the timing and duration of alluvial fan formation. We achieved this in two steps. First, we developed a probabilistic understanding of the age of each AHi subunit by estimating their posterior age distributions in a Bayesian framework (similar to \citealt{MICHAEL2016}). Second, we implemented two statistical models that represent hypothetical formation scenarios for the alluvial fans. Parameter estimation for these models enabled inferences about the timing of alluvial fan formation. 

\subsection{Dating the AHi subunits}

\citet{MICHAEL2016} described the use of Bayesian posteriors for estimating planetary surface ages, under the name “Poisson Timing Analysis.” In this framework, crater counts were assumed to be generated as a Poisson process, with underlying parameter likelihoods, $\mathcal{L}$, given by the standard equation:

\begin{equation}
    \mathcal{L}(t|k,A) = P(k|t,A) = \frac{(A\lambda_t)^k}{k!}e^{-A\lambda_t}
\end{equation}

where $k$ is the observed number of craters, $A$ is the area of the count region, and $\lambda_t$ is the expected density of craters on a surface with age $t$. The correspondence between age and expected density is known as the chronology function, and throughout this study, we adopt the chronology function of \citet{MICHAEL2013} (based on \citet{HARTMANN2005} and \citet{NEUKUM2001}):

\begin{equation}
    \lambda_t(D) = C(D)[(3.79 \times 10^{-14})(e^{6.93t} - 1) + (5.84 \times 10^{-4})t]
\end{equation}

where $\lambda_t(D)$ is the expected density of craters $> D$ km in diameter, and $C(D)$ is a function that accounts for the size-frequency distribution of craters (Table 1 in \citet{MICHAEL2013}). If, a priori, each candidate age is assumed to be equally likely, the likelihood function can be normalized (scaled such that its integral equals 1) and treated as a posterior probability distribution for subunit age. From there, Michael et al. (2016) computed median ages, as well as the $25\%$ and $75\%$ age percentiles. These corresponded to their final age estimates and error bounds, respectively. 

\begin{figure}[h!]
    \plotone{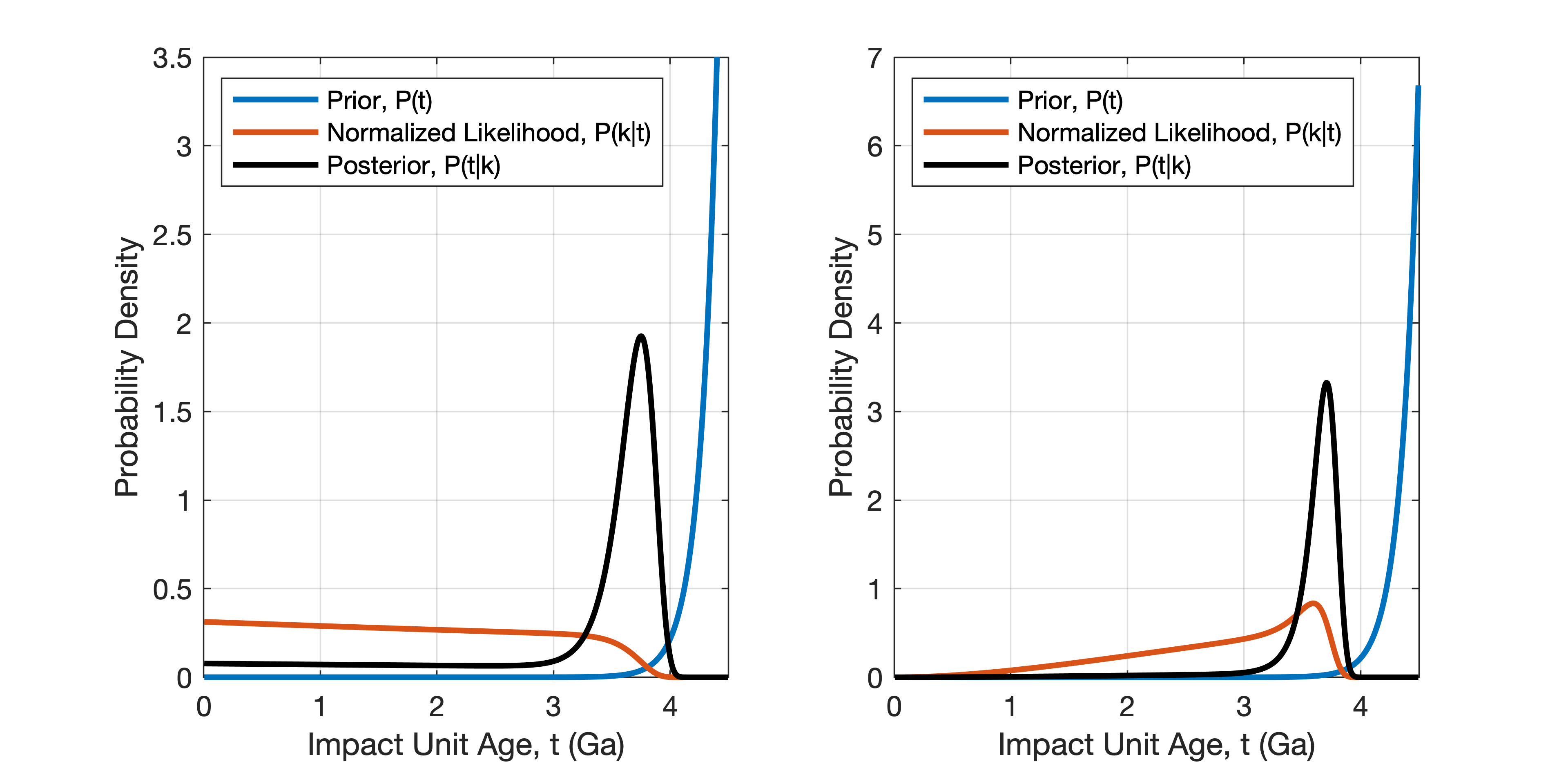}
    \caption{Example normalized likelihood functions, as well as prior and posterior distributions, for the ages of two AHi subunits. For the subunits in the left panel, no craters were counted on the AHi unit ejecta (none $> 4$ km in diameter were found within the count annulus), hence the maximum likelihood is achieved at $t=0$ Ga. For the subunits in the right panel, two craters were counted on the AHi subunit ejecta, hence the maximum likelihood is achieved at $t \sim 3.5$ Ga. Prior distributions are assumed to be the same for each AHi subunit. Despite different likelihood functions, the posterior distributions are similar.}
    \label{fig:fig2}
\end{figure}

In our framework, we had strong reason to believe a non-uniform prior distribution in AHi subunit formation age. In particular, impact fluxes on Mars were elevated $> 3$ Ga, relative to today (see among many others, \citet{MICHAEL2013}). This implies that the AHi subunits, which are themselves formed by impact craters, are more likely to have formed early in Mars’ history. Thus, for our application, we took the prior cumulative distribution function (CDF) of AHi subunit ages to be proportional to the chronology function, truncated at $4.5$ Ga. A sensitivity test truncating at $3.54$ Ga is discussed in \S4.). We then computed the prior probability density function (PDF) on AHi subunit ages as the normalized derivative of the \citet{MICHAEL2013} chronology function (Figure \ref{fig:fig2}):

\begin{equation}
    P(t) \propto \frac{d}{dt} [(3.79 \times 10^{-14})(e^{6.93t} - 1) + (5.84 \times 10^{-4})t].
\end{equation}

Using the likelihood function and prior distribution, we computed posterior age PDF's (Figure \ref{fig:fig2}) by multiplying and normalizing:

\begin{equation}
    P(t|k,A) \propto \mathcal{L}(t|k,A)P(t)
\end{equation}

\begin{figure}[h!]
    \plotone{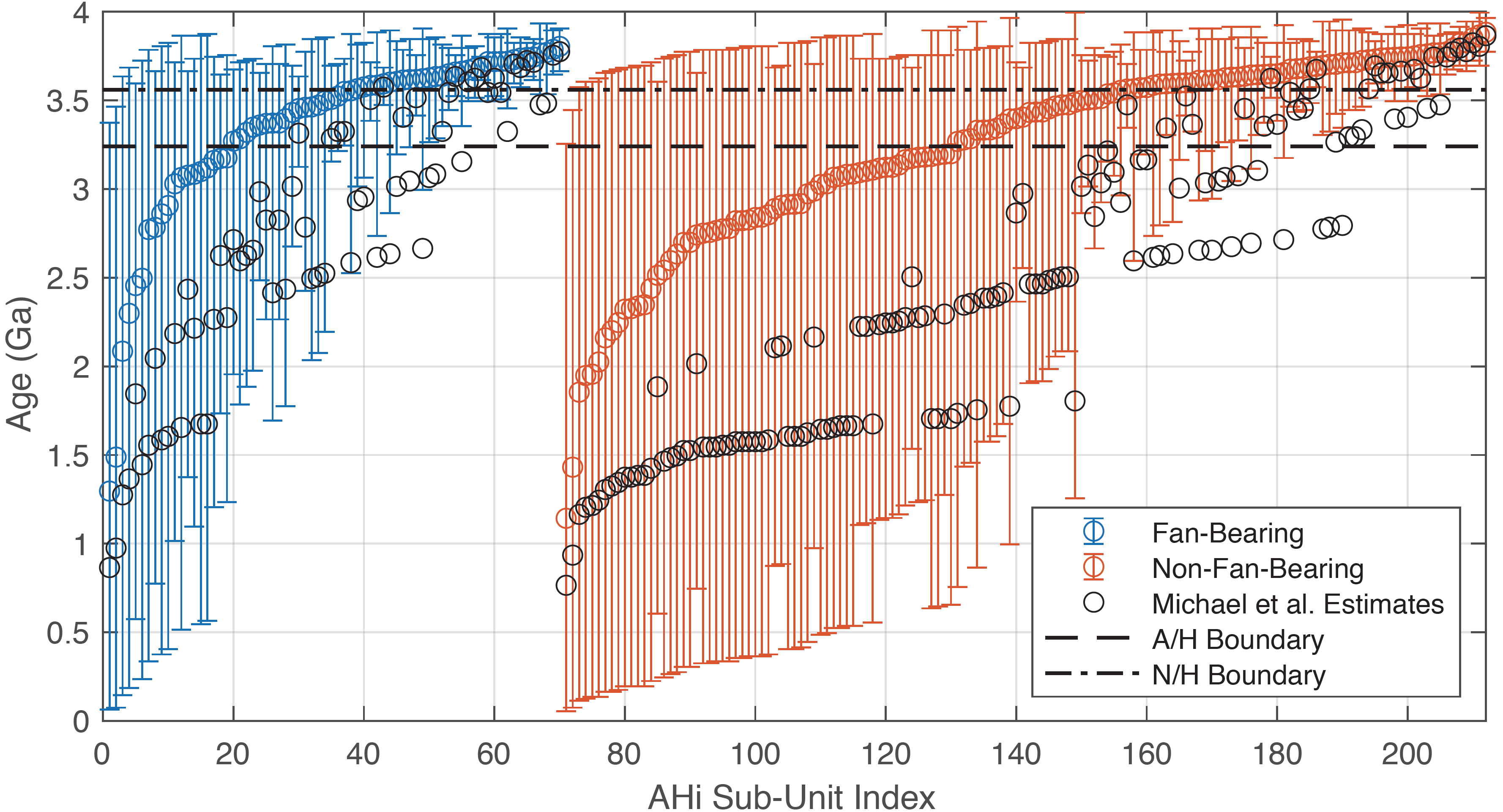}
    \caption{Individual age estimates and $2\sigma$ error bars for each AHi subunit using the Bayesian posterior method. Central estimates for our method are expected ages marginalized over age posterior distributions, while the \citet{MICHAEL2016} method selects median values from the normalized likelihood function. Our strong prior systematically biases our age estimates upward, relative to \citet{MICHAEL2016}.}
    \label{fig:fig3}
\end{figure}

Our estimates are consistent with the method of \citet{MICHAEL2016} in $\sim 90\%$ of cases (Figure \ref{fig:fig3}). In every case, our strong prior forces our central estimate to be older than the central estimate from the \citet{MICHAEL2016} method (Figure \ref{fig:fig3}). Almost all AHi subunits are consistent at the $95\%$ confidence level with being Hesperian or Amazonian in age, however (considering only crater data and setting aside cross-cutting relationships) most of the AHi subunits are also consistent with being Noachian in age \citep[e.g.][]{FASSETT2016}.

\begin{figure}[h!]
    \plotone{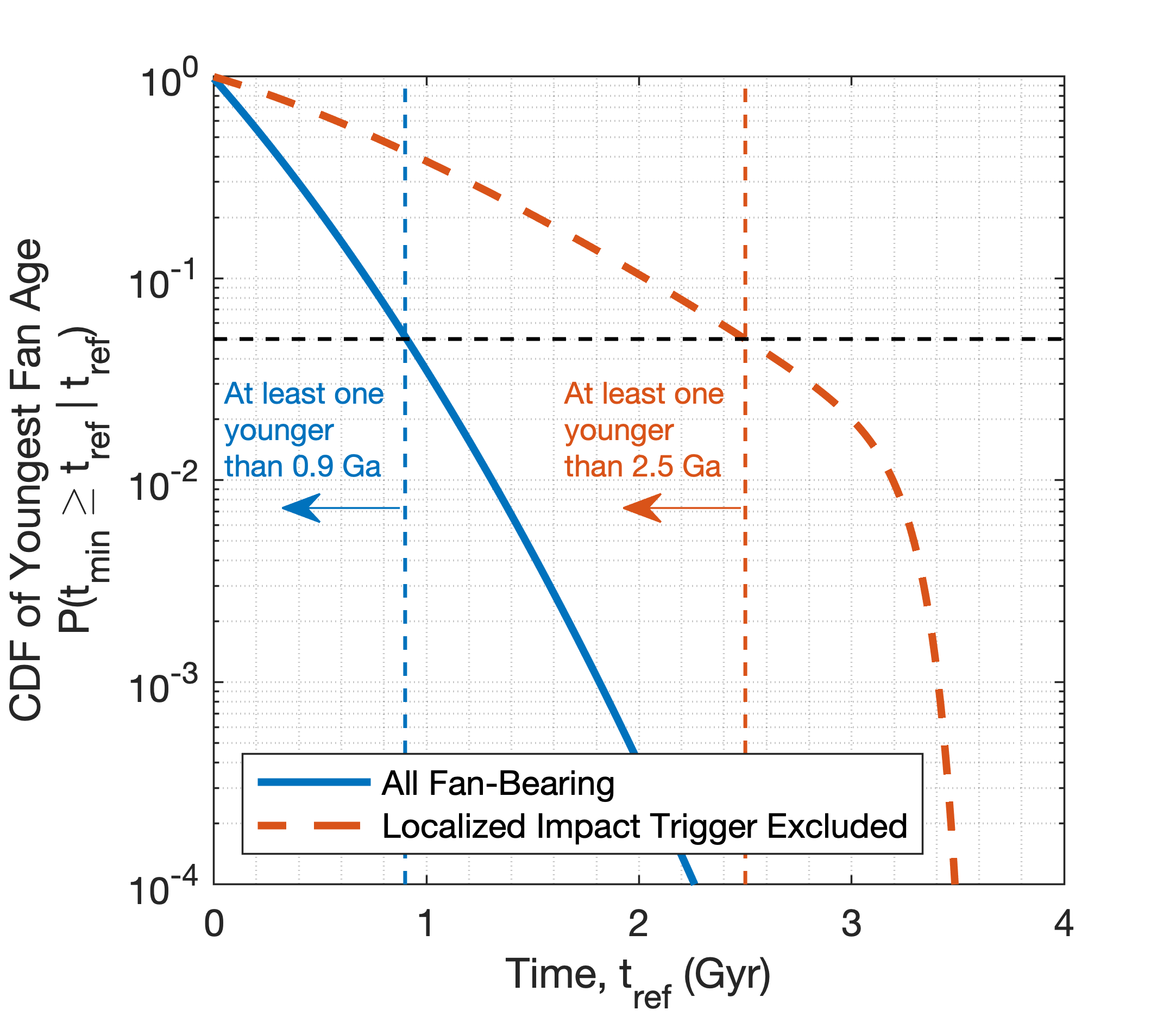}
    \caption{Cumulative distribution functions (CDF's) for the age of the youngest alluvial fan, $t_{min}$, as a function of $t_{ref}$. The blue curve is the CDF for all fan-bearing subunits, while the red curve is the CDF for those fan-bearing subunits where localized impact trigger is excluded. The horizontal, black dashed line indicates the $P = 0.05$ level. Thus, the red and blue vertical dashed lines (defined by the intersection of the CDF's with the $P = 0.05$ level) represent our $95\%$ confidence level upper estimates for the age of the youngest fan in each population. Our results indicate that at least one fan likely formed in the last $\sim 0.9$ Ga and that at least one fan where localized impact trigger is excluded formed in the last $\sim 2.5$ Ga.}
    \label{fig:fig4}
\end{figure}

Using the posterior age CDF's for each AHi subunit, we derived the CDF for the age of the youngest fan-bearing AHi subunit. To do so, we assumed the ages of each AHi subunit are independent random variables: 

\begin{equation}
    P(t_{min} \geq t_{ref}|t_{ref}) = \prod_i P(t_i \geq t_{ref}|k_i,A_i, t_{ref})
\end{equation}

where $i$ is an arbitrary index over the fan-bearing AHi subunits. We computed this distribution for two populations: (a) all fan-bearing AHi subunits (N = 70), and (b) those fan-bearing AHi subunits where localized impact trigger is excluded as an explanation for alluvial fan formation (N = 13). Population (b) comprises those fan-bearing subunits in which pre- or syn-fluvial impact craters were identified \citep{KITE2017,KITE2021B}. The results are summarized in Figure \ref{fig:fig4}. We found at the $95\%$ confidence level that $t_{min}\lesssim 0.9$ Ga if we include all fan-bearing AHi subunits. Considering only fan-bearing AHi subunits where localized impact trigger is excluded, we found at the $95\%$ confidence level that $t_{min}\lesssim 2.5$~Ga.

\subsection{Modeling Fan Formation Scenarios}

\begin{figure}[h!]
    \plotone{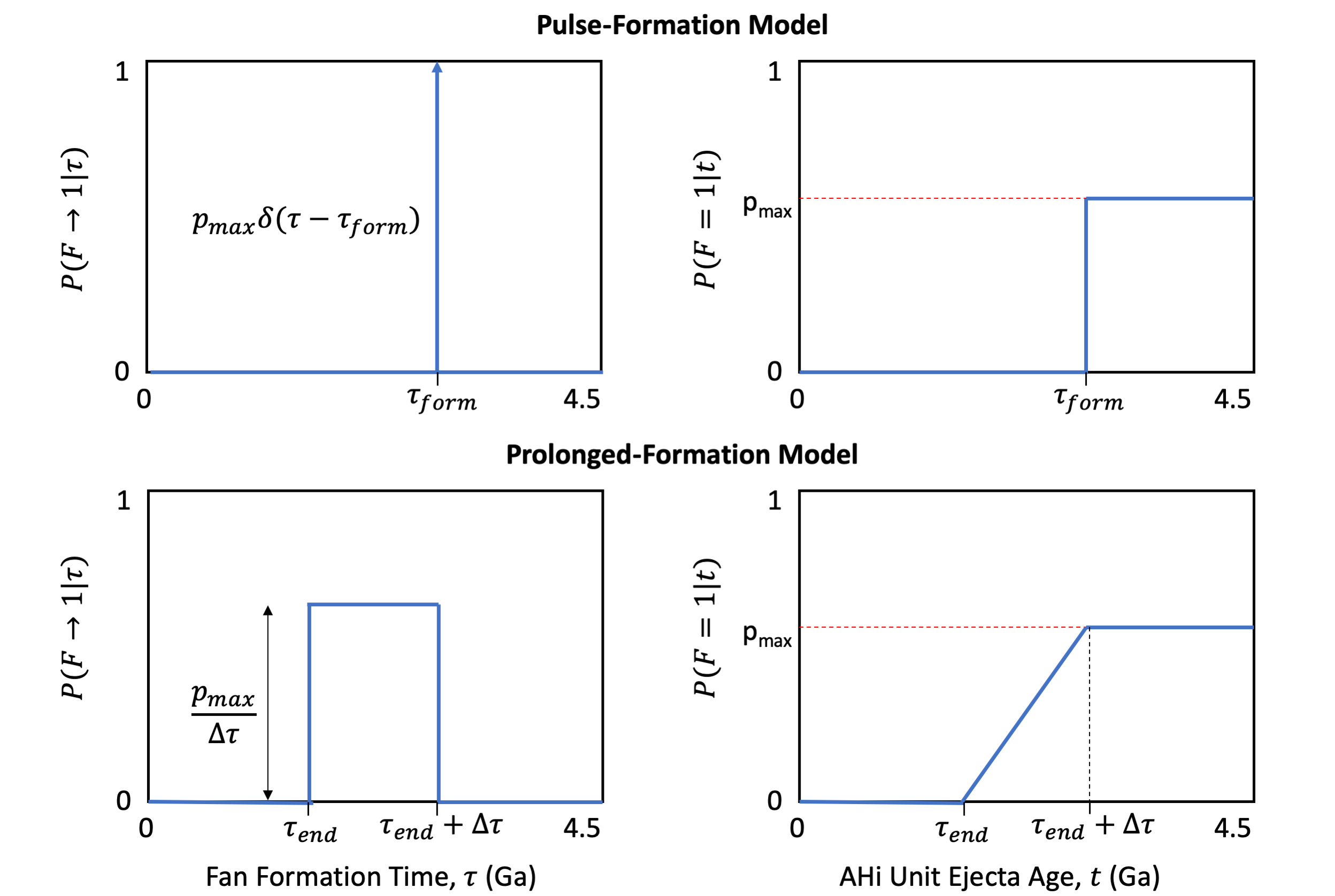}
    \caption{Schematic illustrating the two statistical models for fan formation. The pulse-formation model is on the top row and the prolonged-formation model is on the bottom. The left column for both models shows the instantaneous view, where the probability of an AHi subunit acquiring alluvial fans, $P(F \rightarrow 1)$, is plotted as a function of time, $\tau$. The right column for both models shows the cumulative view, where the probability of an AHi subunit having a fan, $P(F=1|t)$, is shown as a function of AHi subunit age, $t$. Figure is illustrative and not drawn to scale. }
    \label{fig:fig5}
\end{figure}

To investigate the formation timing of alluvial fans in a way that incorporates information from crater-counts on non-fan-bearing AHi subunits, we developed two simple statistical models (Figure \ref{fig:fig5}). In the simplest model (pulse-formation model), we assumed that alluvial fans formed at a single moment in history, $\tau_{form}$, and that all AHi subunits with age $t_i \geq \tau_{form}$, had an equal probability, $p_{max}$, of acquiring alluvial fans. In the second model (prolonged-formation model), each AHi subunit had a constant instantaneous probability, $p_{max}/\Delta\tau$, of acquiring alluvial fans during the $\Delta\tau$-long time period ending at $\tau_{end}$. This prolonged-formation model is approximately equivalent to a model with many short bursts of activity that are evenly spaced or randomly spaced throughout the $\Delta\tau$-long time period ending at $\tau_{end}$. In both models, the probability of an AHi subunit younger than $\tau_{end}$ or $\tau_{form}$ acquiring fans was assumed to be 0. Both models are motivated by searching for changes in the probability of acquiring fans as a function of time, for example due to atmospheric evolution. Neither model is set up to identify preferred spatial patches for fan formation, for example latitude dependent climate variations. Any such variations are averaged out in the globally uniform (but time-varying) probability of acquiring alluvial fans.

For both statistical models, the formation scenarios can be viewed both in an instantaneous framework (left panel, Figure \ref{fig:fig5}), and in a cumulative framework (right panel, Figure \ref{fig:fig5}). Both viewpoints are instructive, as the instantaneous view relates to a hypothetical climate scenario (e.g. short-lived, global transient greenhouse), while the cumulative view relates fan-acquisition probabilities to AHi subunit ages, $t_i$. The relationship between these two views can be formalized mathematically via simple integrals for both models:

\begin{equation}
    \mathcal{L}(\tau_{form},p_{max}|F_i = 1, t_i) = \int_0^{t_i} P(F \rightarrow 1 | \tau_{form}, p_{max}, \tau)d\tau
\end{equation}

\begin{equation}
    \mathcal{L}(\tau_{end},\Delta\tau,p_{max}|F_i = 1, t_i) = \int_0^{t_i} P(F \rightarrow 1 | \tau_{end},\Delta\tau, p_{max}, \tau)d\tau
\end{equation}

where $F_i=1$ when the $i^{th}$ AHi subunit contains fans (0 otherwise), and $P(F \rightarrow 1)$ is the probability of acquiring fans, which varies with time (Figure \ref{fig:fig5}). Note, because $F_i$ is a binary variable, $P(F_i=0)=1- P(F_i=1)$, which defines $\mathcal{L}$ in the case that $F_i=0$.

Because we only know the AHi subunit ages, $t_i$, probabilistically, we computed expected likelihoods (instead of the likelihoods themselves) as a function of model parameters and crater density data by marginalizing over the age posterior distributions:

\begin{equation}
    E(\mathcal{L}(\tau_{form},p_{max}|F_i, k_i,A_i)) = \int_0^{4.5}\mathcal{L}(\tau_{form},p_{max}|F_i, t)P(t|k_i,A_i)dt
\end{equation}

\begin{equation}
    E(\mathcal{L}(\tau_{end},\Delta\tau,p_{max}|F_i, k_i,A_i)) = \int_0^{4.5}\mathcal{L}(\tau_{end},\Delta\tau,p_{max}|F_i, t)P(t|k_i,A_i)dt
\end{equation}

where $k_i$ is the observed crater count and $A_i$ the count region area for the i’th (out of 212) AHi subunits. Taking the data from every AHi subunit together, we can multiply the expected likelihoods together to estimate the overall expected likelihood as a function of model parameters and data: 

\begin{equation}
    E(\mathcal{L}(\tau_{form},p_{max}|\vec{F}, \vec{k},\vec{A})) = \prod_i E(\mathcal{L}(\tau_{form},p_{max}|F_i, k_i,A_i))
\end{equation}

\begin{equation}
    E(\mathcal{L}(\tau_{end},\Delta\tau,p_{max}|\vec{F}, \vec{k},\vec{A})) = \prod_i E(\mathcal{L}(\tau_{end},\Delta\tau,p_{max}|F_i, k_i,A_i))
\end{equation}

\begin{figure}[h!]
    \plotone{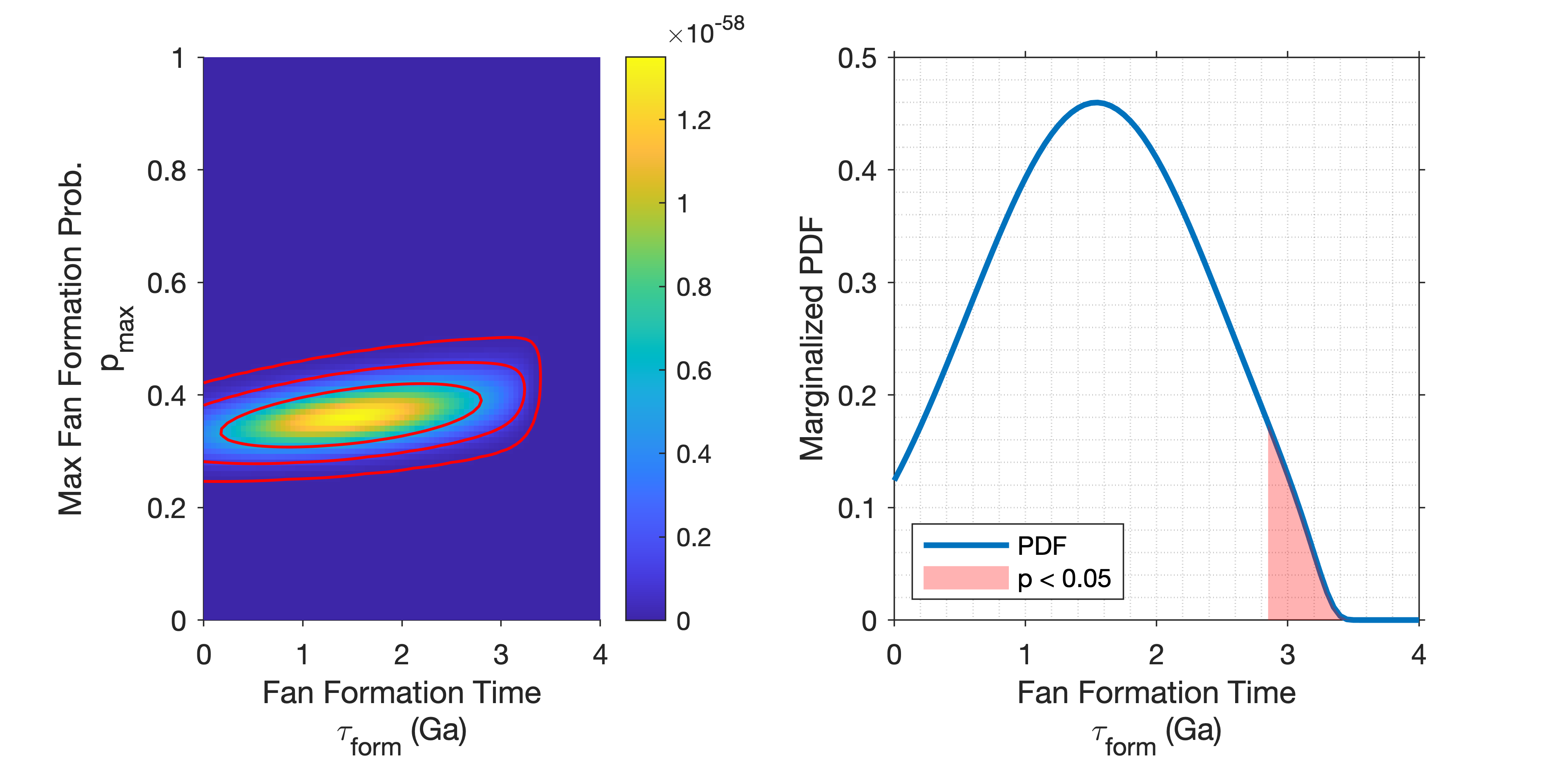}
    \caption{Left: expected likelihood for the pulse-formation model, as a function of $\tau_{form}$ and $p_{max}$ (from Equation 10). Red contours show 1-, 2-, and 3-$\sigma$ confidence regions (68, 95, and 99.7 percentile regions represented by inner, middle, and outer red contours, respectively). Right: marginalized PDF for $\tau_{form}$, assuming a uniform prior over $p_{max}$.}
    \label{fig:fig6}
\end{figure}

We computed the expected likelihood for the pulse-formation model as a function of $\tau_{form}$ and $p_{max}$, allowing them to vary in [0 Ga, 4 Ga] and [0,1], respectively (Figure \ref{fig:fig6}). The model achieved maximum likelihood at $\tau_{form}=1.5$ Ga and $p_{max} = 0.36$. Assuming a uniform prior over $p_{max}$, we marginalized the likelihoods over $p_{max}$ (``summed over $p_{max}$'') to estimate a posterior PDF for $\tau_{form}$ (right panel of Figure \ref{fig:fig6}). The $95\%$ confidence level for the estimated posterior PDF in the right panel of Figure \ref{fig:fig6} is slightly more restrictive than the right-most extent of the middle red contour in the right panel of Figure \ref{fig:fig6} - this is due to the marginalization over $p_{max}$. At the $95\%$ confidence level, we found that in the pulse-formation scenario, the data requires fan formation to occur within the last $\sim 2.8$ Gyr. 

\begin{figure}[h!]
    \plotone{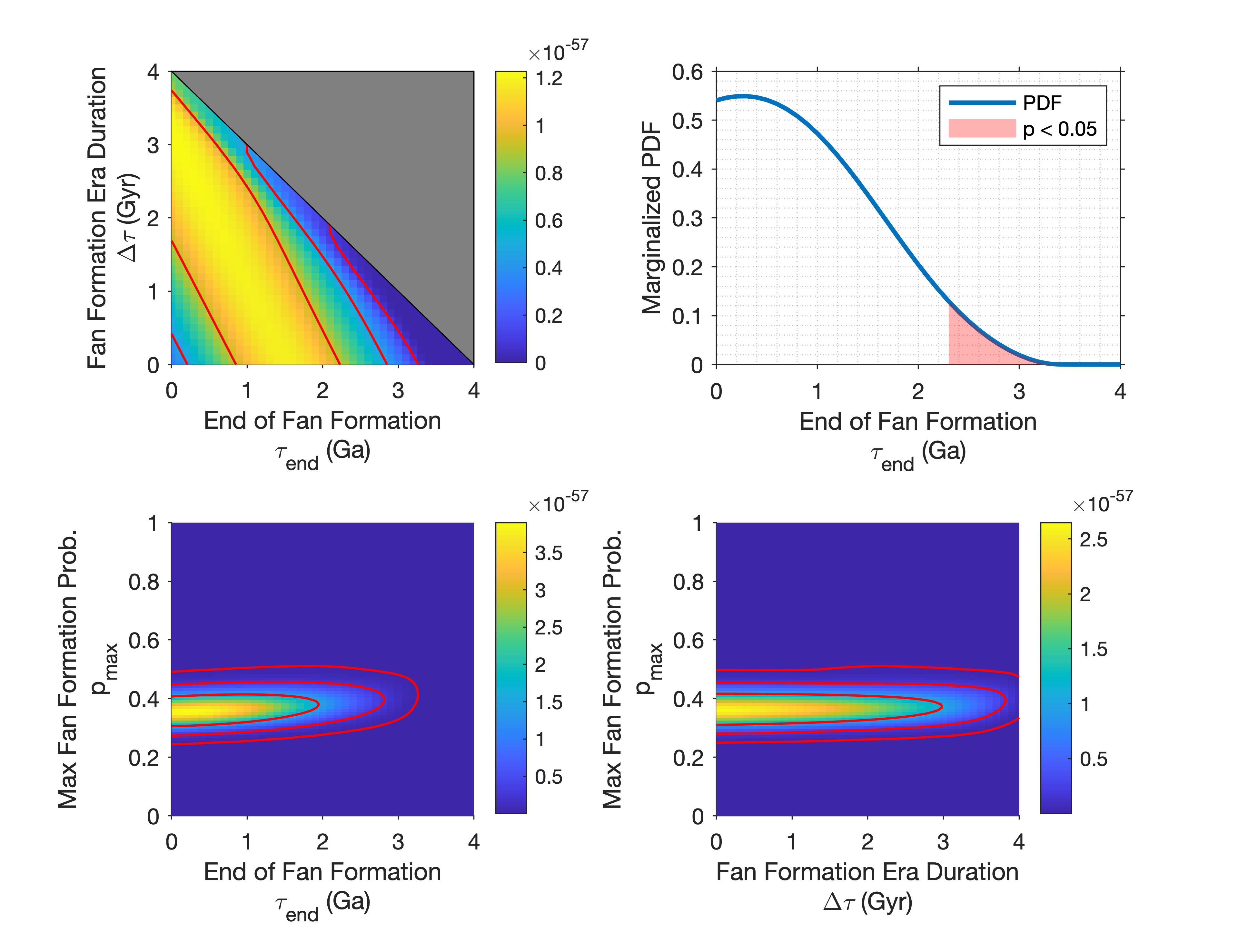}
    \caption{Top left: expected likelihood for the prolonged-formation model, as a function of  $\tau_{end}$ and $\Delta\tau$, marginalizing over $p_{max}$. Greyed-out region was not modeled. Bottom left: expected likelihood for the prolonged-formation model, as a function of $\tau_{end}$ and $p_{max}$, marginalizing over $\Delta\tau$. Top right: estimated posterior distribution for $\tau_{end}$, marginalizing over $\Delta\tau$ and $p_{max}$. Bottom right: expected likelihood for the prolonged-formation model, as a function of $p_{max}$ and $\Delta\tau$, marginalizing over $\tau_{end}$. Red contours show 1-, 2-, and 3- $\sigma$ confidence regions in the marginalized, 2-D parameter spaces. Note, this is not equivalent to the projection of confidence regions in 3-D parameter space down to 2-D. The apparent simultaneous maxima of $\tau_{end} = 0$ and $\Delta\tau = 0$ in the bottom panels arise from the fact that we restricted the domains of $\tau_{end}$ and $\Delta\tau$ (greyed out region in top-left panel) and assumed uniform prior distributions over those domains. The bottom panels do, however, show that $p_{max}$ is strongly determined by the data and can safely be marginalized out of the inferred posterior distribution (i.e. one can interpret the top-left panel without too much concern over the effect of prior distribution choices). Expected likelihoods are calculated using Equation 11.}
    \label{fig:fig7}
\end{figure}

We computed the expected likelihood for the prolonged-formation model, as a function of $\tau_{end}$, $\Delta\tau$, and $p_{max}$, allowing them to vary in [0 Ga, 4 Ga], [0, 4 Ga $-  \tau_{end}$], and [0,1], respectively (Figure \ref{fig:fig7}). The model achieves maximum likelihood at $\tau_{end}= 0$ Ga, $\Delta\tau=3$~Ga, $p_{max}=0.36$ (Figure \ref{fig:fig7}, top left panel). We compared the maximum likelihoods from the two models (this comparison does not require assuming that either model is a good model). Despite a maximum likelihood solution with $\Delta\tau>0$, the difference in maximum likelihoods between the prolonged-formation model and the pulse-formation model (equivalent to the prolonged-formation model when $\Delta\tau=0$) was not sufficiently large to reject the pulse-formation model via a likelihood-ratio test ($p < 0.81$, where our computed test statistic, $\lambda_{LR} \approx 0.06$, is given by twice the difference in the maximum log-likelihoods for each model and is asymptotically chi-square distributed with 1 degree of freedom). Assuming uniform priors, we marginalized the likelihoods over $\Delta\tau$ and $p_{max}$ to estimate a posterior distribution for $\tau_{end}$ (Figure \ref{fig:fig7}). At the $95\%$ confidence level, we found that in the prolonged-formation scenario, our dataset requires fan formation to have persisted into the last $\sim 2.3$~Gyr.

\section{Discussion}

Using the crater counts on AHi subunits’ ejecta blankets, as well as some simple statistical models, we were able to place constraints on the timing of the formation of alluvial fans. Our estimate of the CDF for the age of the youngest fan indicated that at least one fan likely formed by a process other than localized impact trigger in the last $\sim 2.5$~Gyr. Further, our CDF estimate indicates that at least one fan likely formed in the last $\sim 0.9$ Gyr. This is broadly consistent with the results from the pulse-formation model (which incorporates information from crater-counts on non-fan-bearing AHi subunits), in which the best-fitting time of fan formation was $\sim 1.5$ Gyr. This indicates that fan formation likely persisted well into the Amazonian epoch, consistent with \citet{PALUCIS2020} and \citet{GRANT2019}. Although the pulse-formation model does not favor formation occurring in the last $\sim 1$ Gyr (due to  the fact that very young values of $\tau_{form}$ force non-fan-bearing AHi subunits to take on likelihood $1-p_{max}$ instead of 1 with too high a probability), it does not explicitly reject it. Further, at least some of the young fans can be explained by localized-impact trigger \citep[e.g.][]{WILLIAMS2008, GODDARD2014}, which is not permitted in the pulse-formation model scenario.  

We estimated parameters for a more complex model, the prolonged-formation model, and compared the results with the pulse-formation model. The introduction of a third parameter introduced a large degeneracy between $\tau_{end}$ and $\Delta\tau$ in our parameter retrieval (Figure \ref{fig:fig7}). Although the model selected $\tau_{end}=0$ Ga with fan-formation lasting several Gyr, a likelihood ratio test revealed that the maximum likelihood is not sufficiently higher in the prolonged-formation model than in the pulse-formation model to definitively justify the addition of the third parameter. Thus we cannot reject either the pulse-formation or prolonged-formation model. Further, we do not have tight constraints on $\tau_{end}+\Delta\tau$, due to the degeneracy in our prolonged-formation model parameter retrieval. As a result, we cannot conclusively say, using this crater data alone, whether the alluvial fans formed in an era entirely separate from the valley networks (i.e. $\tau_{end}+\Delta\tau \lesssim 3.4$~Ga with high confidence), or if the alluvial fans represent a long tail of continuous fluvial activity (i.e. $\tau_{end}+\Delta\tau \gtrsim 3.4$~Ga with high confidence). 

For both models, we showed marginalized distributions for $\tau_{form}$ and $\tau_{end}$, assuming uniform priors over the other parameters, $p_{max}$ and $\Delta\tau$. Unlike the prior distributions for the ages of the AHi subunits, we had no particular reason to choose these parameter prior distributions (although we note that marginalizing over $p_{max}$ is likely okay due to its strong determination by the data, see Figure \ref{fig:fig7}). The marginalized distributions should thus be viewed only as condensed data visualizations, rather than true posterior distributions. Despite this, maximum likelihood estimates from both models indicate fan formation continued into the last $\sim 1.5$ Gyr. This global age is substantially ($\sim 2$ Gyr) younger than existing quantitative estimates of the age of fan formation from a study of two regions \citep{MANGOLD2012CHRON}. This can be explained by the fact that our model allows for spatially patchy fan formation (i.e. $p_{max} < 1$) while the method of \citet{MANGOLD2012CHRON} assumes uniform fan formation and thus would overestimate the young limit on alluvial fan formation if applied to our dataset. We note that the assumption that $p_{max}=1$ may be perfectly acceptable within the \citet{MANGOLD2012CHRON} study areas. Our results are consistent with young age estimates obtained using smaller count areas for some fans (e.g. \citealt{GRANT2011, GRANT2019}), and are also consistent with Amazonian age estimates made for some non-fan features that indicate flowing surface liquid water (e.g., \citealt{DICKSON2009,FASSETT2010,HOWARDMOORE2011,WEITZ2013, EHLMANNBUZ2015,BROSSIER2021,BUTCHER2021}). A K-Ar age of (2.12 $\pm$ 0.36) Ga obtained for aqueous minerals at Gale crater by the \emph{Curiosity} rover \citep{MARTIN2017} is consistent with our scenario but does not require it. Our strategy in this study does not incorporate the prior indications of young activity from these previous studies, and our crater data does not overlap with that used by these previous studies, therefore our results are independent.

By themselves, our data do not allow us to determine whether fan formation was intermittent (either synchronous between fan-forming regions, or regionally diachronous), or alternatively if water flow occurred in more-or-less every year in all regions where fans are present. Consensus on this important topic may have to await a future landed mission.

\begin{figure}[h!]
    \plotone{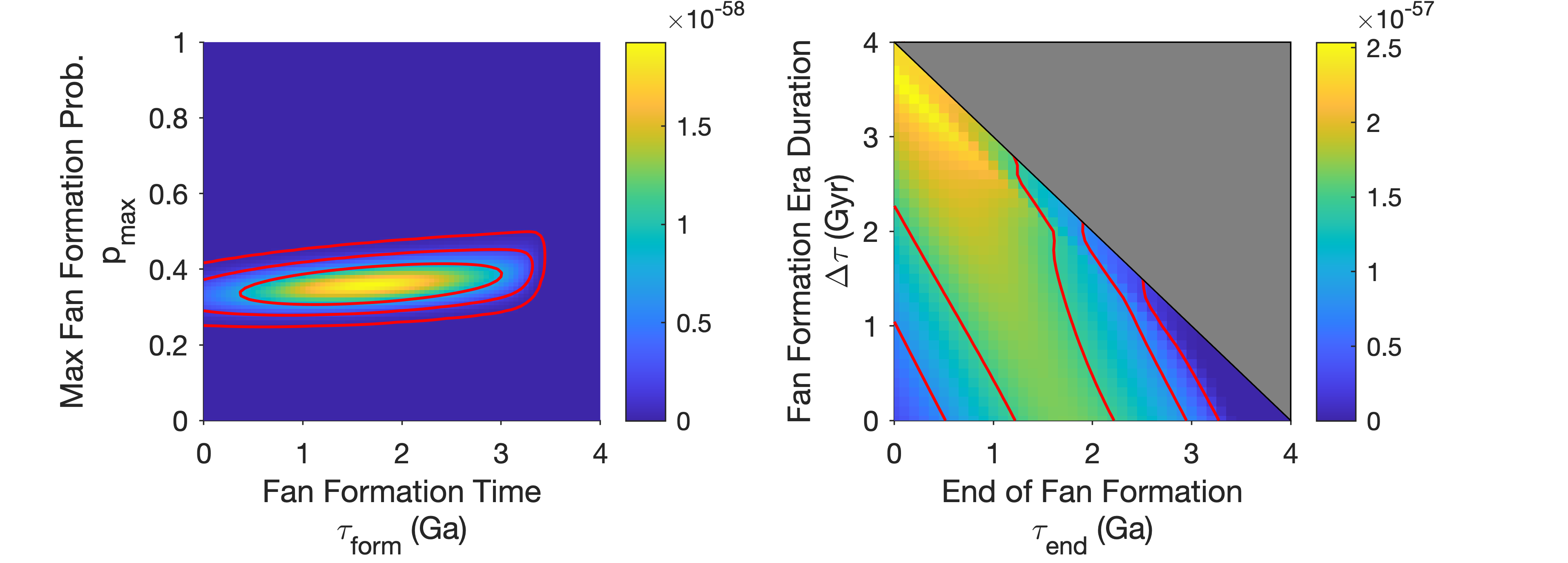}
    \caption{Model retrievals with the outer annulus limit of our count area removed. Left: expected likelihood for the pulse-formation model, as a function of $\tau_{form}$ and $p_{max}$.  The results are biased to older ages relative to those in Figure 4.6, as expected given increased contamination by “poke-through” craters. Right: expected likelihood for the prolonged-formation model, as a function of  $\tau_{end}$ and $\Delta\tau$, marginalizing over $p_{max}$. Greyed-out region was not modeled. Red contours show 1-, 2-, and 3-$\sigma$ confidence regions in both panels (constructed in marginalized, 2-D space in the right panel).}
    \label{fig:fig8}
\end{figure}

In general, data availability limited our parameter retrievals for $\tau_{form}$ and $\tau_{end}$, which  had large uncertainties associated with them. Indeed, our crater counts were sparse due to our lower diameter cutoff of 4 km, causing our AHi subunit age posteriors to be largely influenced by our choice of prior distribution and have high variance (Figures \ref{fig:fig2} and \ref{fig:fig3}). We performed a series of sensitivity tests to illustrate how the model performs both with more data and with idealized data. First, we performed the parameter estimation with the same data, but with the outer annulus cutoff removed (Figure \ref{fig:fig8}). This had the effect of biasing our retrievals to older ages due to the larger influence of old craters “poking-through” the ejecta blankets near the edges (Figure \ref{fig:fig8}). Indeed, the pulse-formation model achieved maximum likelihood at $\tau_{form} = 1.7$~Ga and $p_{max} = 0.36$. Further, the increased quantity of data decreased the variance of our estimates, which is evident in the decreased degeneracy in the prolonged-formation model (Figure \ref{fig:fig8}). 

\begin{figure}[h!]
    \plotone{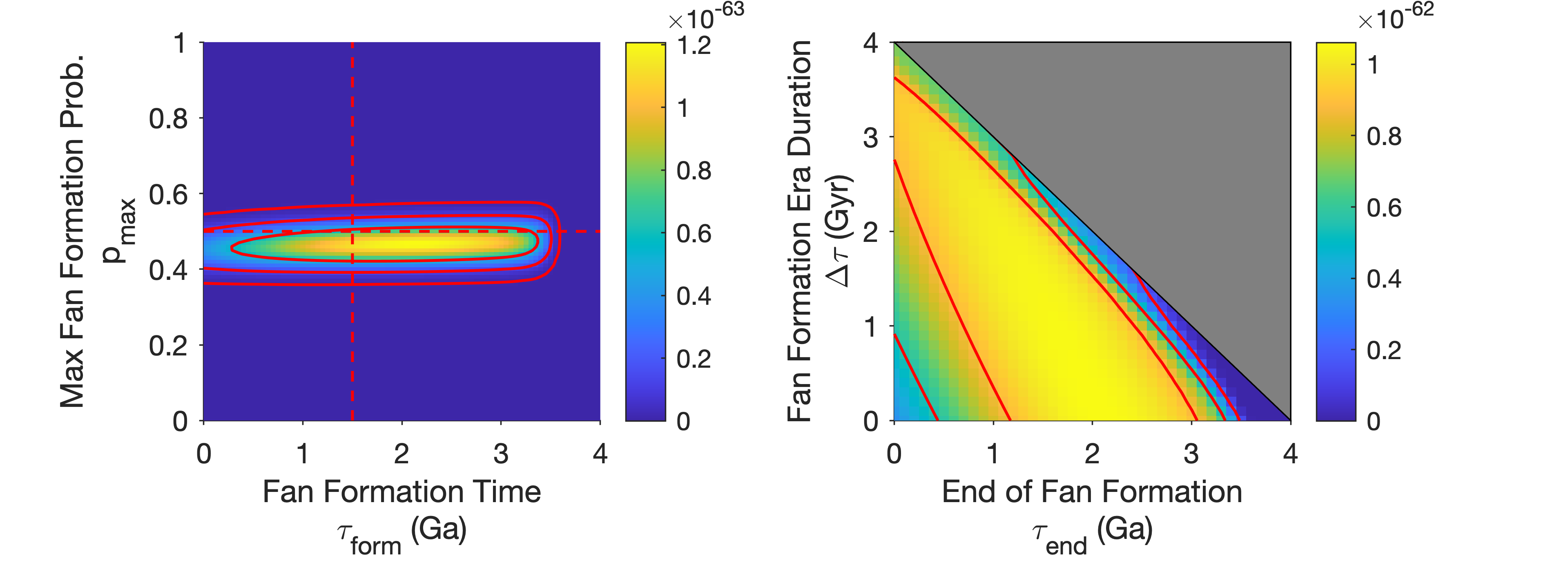}
    \caption{Model retrievals where synthetic data was created assuming the pulse-formation model is correct (i.e. $\Delta\tau = 0$). Left: expected likelihood for the pulse-formation model, as a function of $\tau_{form}$ and $p_{max}$. Dashed red lines show true parameters from the synthetic data generation process. Right: expected likelihood for the prolonged-formation model, as a function of  $\tau_{end}$ and $\Delta\tau$, marginalizing over $p_{max}$. Greyed-out region was not modeled. Red contours show 1-, 2-, and 3-$\sigma$ confidence regions in both panels (constructed in marginalized, 2-D space in the right panel).}
    \label{fig:fig9}
\end{figure}

Next, in order to evaluate how effectively the mathematical model can retrieve parameter values, we carried out a test using synthetic data. Specifically, we performed the parameter retrievals using the original count areas (including outer annulus limit),  using synthetic data where the absence or presence of alluvial fans and ejecta crater counts were randomly generated assuming that the pulse-formation model was true, that the fans formed at 1.5 Ga with probability 0.5, and that AHi subunits have ages drawn from the prior distribution, truncated at 4 Ga. In general, the model performs well and achieves maximum likelihood within $1\sigma$ of the correct value (Figure \ref{fig:fig9}). However, there is still substantial variance in the results, and the model is unable to substantially break the degeneracy between the end of fan formation and the duration of the fan formation era, although it does achieve maximum likelihood at $\tau_{end} = 2.1$ Ga, $p_{max} = 0.47$  and $\Delta\tau = 100$~Myr (Figure \ref{fig:fig9}). This suggests that more craters are needed to constrain the formation timing.

\begin{figure}[h!]
    \plotone{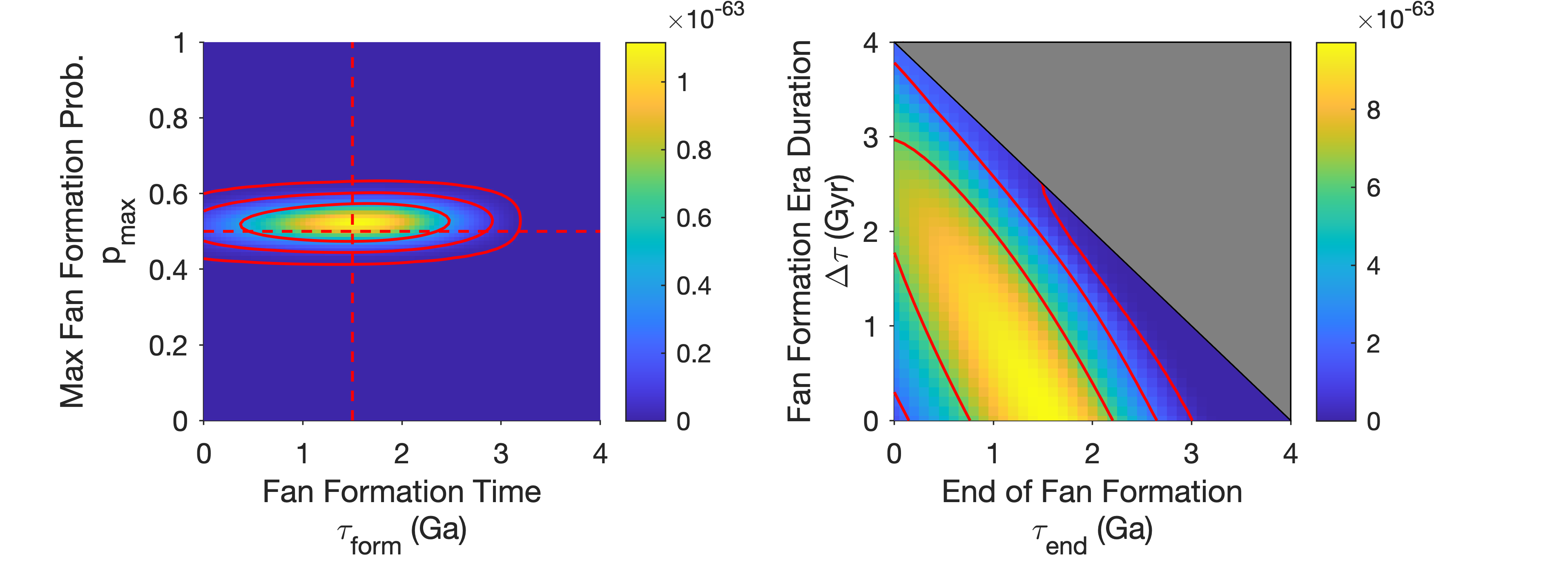}
    \caption{Model retrievals where synthetic data was created assuming the pulse-formation model is correct (i.e. $\Delta\tau = 0$) and inflating the count area by a factor of 10. Left: expected likelihood for the pulse-formation model, as a function of $\tau_{form}$ and $p_{max}$. Dashed red lines show true parameters from the synthetic data generation process. Right: expected likelihood for the prolonged-formation model, as a function of  $\tau_{end}$ and $\Delta\tau$, marginalizing over $p_{max}$. Greyed-out region was not modeled.  Inflating the count area had the effect of decreasing the variance in our estimates. Red contours show 1-, 2-, and 3-$\sigma$ confidence regions in both panels (constructed in marginalized, 2-D space in the right panel).}
    \label{fig:fig10}
\end{figure}

We repeated the experiment with synthetic data, artificially increasing the AHi count areas by a factor of 10. This had the effect of placing roughly 10 times as many craters on each subunit, thereby decreasing the uncertainty in the posterior age distributions. As a result, the variance in our parameter retrievals shrunk (compare Figure \ref{fig:fig10} with Figure \ref{fig:fig9}). Further, the parameter retrieval has a slightly increased ability to discriminate between the prolonged- and pulse-formation models (relative to the synthetic data experiment with uninflated AHi subunit areas) and preferentially selects a fan formation era duration of 0 (Figure \ref{fig:fig10}). This indicates that in order to obtain better estimates for the timing and duration of alluvial fan formation, one needs substantially more data and must consider craters at smaller diameters (recall our lower cutoff was 4 km). Although we ignored the possibility that surface processes efficiently obliterated 4 km diameter craters in the post-Noachian, one must correct for crater obliteration to avoid underestimating age when smaller craters are considered \citep[e.g.][]{PALUCIS2020}.

In our methodology, we did not account for uncertainty in the chronology function, which is continually being re-evaluated from a combination of crater counting, geochronology, and solar system dynamics perspectives \citep[e.g.][]{WERNERTANAKA2011,ROBBINS2014,FARLEY2014,MARCHI2021}. It is possible that our age estimates are systematically younger or older than the true ages, which would have biased our fan formation timing estimates. This potential effect is especially important in the data-limited case, where the prior strongly informs posterior age distributions (e.g. if there were ample data, the posterior curves in Figure \ref{fig:fig2} would coincide with the normalized likelihood curves). A related point is that if we had imposed an exponentially decaying prior likelihood on fan formation time, then our fan age estimates would have been older. Further, we assumed a maximum age of 4.5 Ga in our prior age distribution. In principle, one could impose a younger maximum age based on stratigraphy to obtain tighter constraints. We re-performed our parameter retrieval assuming a maximum age of 3.54 Ga (approximately the Noachian/Hesperian boundary). However, roughly half of the AHi subunits achieved maximum posterior probability at 3.54 Ga, indicating that this prior was too restrictive. It is possible that with improved crater surveys, the data would produce maximum a-posteriori ages that are well into the Amazonian. 

Finally, we did not account for the possibility that the fan catalog is incomplete. It is possible that some fans were wrongly identified and that other fans were missed in certain craters in the survey. To explore this effect, we performed a sensitivity test where the probability of an AHi subunit acquiring fans after the end of fan formation (or after the pulse of fan formation) was 0.001, instead of 0. This had a negligible effect on the outcome, and we concluded that survey incompleteness needs to be more severe (e.g. a $10\%$ chance of missing all the fans in a crater) to substantially alter the results. 

\section{Conclusions}
We combined a global geologic map \citep{TANAKA2014}, a global database of Martian impact craters \citep{ROBBINS2012}, a global database of alluvial fans \citep{WILSON2021}, and statistical models to make inferences about the timing and duration of alluvial fan formation. Although differing in detail due to differing assumptions, the results of all our models suggest that fan formation had to have persisted into the last $\sim 2.5$ Gyr (based on our most robust and conservative estimate, see Figure \ref{fig:fig4}), well into the Amazonian time period. This estimate is $\sim 1$ Gyr younger than previous estimates \citep[e.g.][]{MANGOLD2012CHRON}. Overall, our modeling procedure was not able to constrain the duration of fan formation. As a result, we were unable to determine from this dataset whether the alluvial fans formed in a long tail of continuous fluvial activity that overlapped with valley-network formation, or if they formed in an era entirely separate from the valley networks.

In general, our inferences were limited by a lack of crater count data, and our tests with synthetic data suggest that detailed surveys of smaller craters on AHi subunit ejecta blankets will enable much tighter constraints on fan formation. Going forward, we hypothesize that more data and more complex statistical models (i.e. ones with more parameters) for the population of alluvial fans (or other geomorphic features) will enable determination of both (a) their joint temporal and spatial evolution and (b) the relative importance of climate-triggered and localized impact-triggered fan formation. The method we have developed can also be applied to determining the ages of other crater-hosted features on Mars. Unraveling the evolution of alluvial fan (and/or other crater-hosted fluvial features') formation would thereby provide even stronger constraints on the evolution of Mars’ climate.



\begin{figure}
\includegraphics[width=\textwidth]{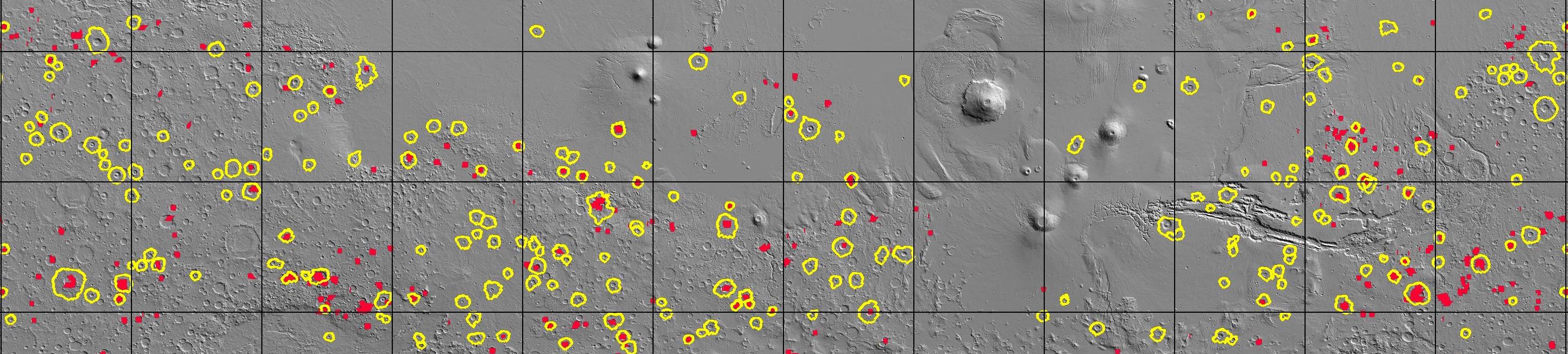}
\caption{\textbf{Appendix Figure.} Map showing the distribution of AHi subunits used in this study (outlined in yellow, from \citet{TANAKA2014}), as well as the distribution of alluvial fans/deltas (red, from \citealt{WILSON2021}).  Latitude/longitude grid spacing is 30$^\circ$. Study region extends from 40$^\circ$S  - 40$^\circ$N and from 0-360$^\circ$E. Background is Mars Orbiter Laser Altimeter shaded relief.}
\end{figure}

\acknowledgments

We thank Stuart Robbins and Marisa Palucis for insightful discussions that helped conceive this study. We thank two reviewers for useful comments. We acknowledge funding from NASA grants 80NSSC20K0144, NNX16AJ38G, and NNX15AM49G.

\bibliography{holo_et_al_revision}{}
\bibliographystyle{aasjournal}

\end{document}